\documentclass[preprint2,trackchanges]{aastex701}

\newcommand\aastex{AAS\TeX}

\usepackage{subfigure}

\usepackage{longtable, booktabs, siunitx}
\begin{document}

\title{Template \aastex v7.0.1 Article with Examples\footnote{Footnotes can be added to titles}}

\title{Revisiting the Neutrino Emission and Physical Parameter Constraints from Time-Resolved Spectra of Gamma-Ray Bursts}

\correspondingauthor{Hou-Jun L\"{u} }
\email{lhj@gxu.edu.cn}

\author[0009-0007-5062-3221]{Yang-Dong-Jun Ou}
\affiliation{Guangxi Key Laboratory for Relativistic Astrophysics, School of Physical Science and Technology, Guangxi University, Nanning 530004, People's Republic of China}
\email{2207301110@st.gxu.edu.cn}

\author[0000-0001-6396-9386]{Hou-Jun L\"{u} }
\affiliation{Guangxi Key Laboratory for Relativistic Astrophysics, School of Physical Science and Technology, Guangxi University, Nanning 530004, People's Republic of China}
\email{lhj@gxu.edu.cn}

\author[0009-0005-3856-7576]{Hao-Yu Yuan}
\affiliation{College of Physics and Electronic Information Engineering, Guangxi Minzu Normal University, Chongzuo 532200, China}
\email{d202280064@alumni.hust.edu.cn}

\author[0009-0007-2424-6469]{Si-Xue Ren }
\affiliation{Guangxi Key Laboratory for Relativistic Astrophysics, School of Physical Science and Technology, Guangxi University, Nanning 530004, People's Republic of China}
\email{r2734723538@gmail.com}

\author[0000-0001-8217-8889]{Xue-Zhao Chang }
\affiliation{Institute of Astrophysics, Chuxiong Normal College, Chuxiong 675000, China}
\email{cxzttwl@163.com}

\author[0000-0001-5348-7033]{Yan-Qiu Zhang}
\affiliation{School of Physics and Electronic Science, Guizhou Normal University, Guiyang 550001, People’s Republic of China}
\email{yanqiuzhang@gznu.edu.cn}

\author[0000-0002-4771-7653]{Shao-Lin Xiong}
\affiliation{State Key Laboratory of Particle Astrophysics, Institute of High Energy Physics, Chinese Academy of Sciences, 19B Yuquan Road, Beijing 100049, People’s Republic of China}
\affiliation{University of Chinese Academy of Sciences, Beijing 100049, People’s Republic of China}
\email{xiongsl@ihep.ac.cn}

\author[0000-0002-7044-733X]{En-Wei Liang }
\affiliation{Guangxi Key Laboratory for Relativistic Astrophysics, School of Physical Science and Technology, Guangxi University, Nanning 530004, People's Republic of China}
\email{lew@gxu.edu.cn}

\begin{abstract}

The non-detection of neutrinos associated with gamma-ray bursts (GRBs) has placed significant constraints on the physical parameter space of GRB jet models. However, most of these constraints have been derived using the time-integrated spectra. In this paper, we adopt time-resolved spectra, which provide a more detailed description of the prompt emission spectral structure, to calculate the neutrino emission, and investigate how they affect the constraints on GRB physical parameters. By selecting three GRBs (i.e., GRBs 090902B, 110721A and 080916C) whose spectra are composed of different components (e.g., thermal, thermal + non-thermal, and non-thermal), we find that the expected neutrino event numbers from the summed time-resolved spectra are lower than those from the time-integrated approach in the region of relatively low dissipation radius $R$ and bulk Lorentz factor $\Gamma$, while the opposite trend is found at high $R$ and high $\Gamma$. Moreover, we also use GECAM-C data to calculate the neutrino emission of the brightest GRB 221009A, and find that the expected number of neutrino events calculated from the summed time-resolved spectra is approximately three times that from the time-integrated spectra in the high-$R$ and high-$\Gamma$ regime. More details of constraints on the physical parameters are also discussed. These results highlight the importance of time-resolved spectra for future GRB neutrino studies.

 
\end{abstract}



\section{Introduction}
High-energy neutrino emission from gamma-ray bursts (GRBs), produced during both the prompt emission phase \citep[e.g.,][]{1997PhRvL..78.2292W,2006PhRvD..73f3002M,2008PhRvD..78b3005M,2009ApJ...691L..67W,2012PhRvD..85b7301L,2013PhRvL.110l1101Z,2017ApJ...843...17X,2022ApJ...941L..10M,2023ApJ...944..115A,2023ApJ...958..113A,2024ApJ...976..174O,2025ApJ...994...91O,2025ApJ...987...79L,2025ApJ...985....9Z,2026ApJ..1010..234Y} and afterglow phase \citep[e.g.,][]{2000ApJ...541..707W,2001ApJ...551..249D,2002A&A...396..303L,2007PhRvD..76l3001M,2006PhRvL..97e1101M,2017PhRvD..96j3004T,2026A&A...710A.168A,2026JHEAp..5300636M}, has been extensively studied over the past decades. Numerous studies have adopted data from different detectors and performed dedicated analyses to search for high-energy neutrino events which are possibly associated with GRBs, but have not detected any such events. \citep[e.g.,][]{2011PhRvL.106n1101A,2015ApJ...805L...5A,2017ApJ...843..112A,2021MNRAS.500.5614A,2024ApJ...964..126A,2024JCAP...08..006A}. Nevertheless, the results of these non-detections of any neutrino events are also invoked to constrain the parameter space of the GRB mechanism \citep[e.g.,][]{2013ApJ...772L...4G,2015ApJ...805L...5A,2016ApJ...824..115A,2017ApJ...843..112A,2022ApJ...941L..10M,2023ApJ...944..115A,2026APh...17803226P,2026ApJ..1006..201V}. However, most of these constraints have been derived using the time-integrated photon spectra, which are essentially an average over the entire burst duration. If this is the case, the time-integrated spectra cannot reveal their behavior of evolution. Such constraints, inferred from averaged spectra, may therefore not accurately reflect the physical conditions of the emission site. For most bright GRBs, a more appropriate approach to study GRBs is to adopt the time-resolved spectrum due to enough photon counts in a small time bin to allow a reasonable spectral fit\footnote{For faint GRBs, the time‑integrated photon spectrum is inevitable due to a limited number of photons.}. The brighter a GRB is, the more suitable it is for a time-resolved spectral analysis. On the other hand, even when the time-integrated spectrum is used, only the brightest events currently provide meaningful constraints on the physical parameter space with existing neutrino detectors. Brighter GRBs are therefore the ideal candidates for constraining the physical parameter space through the existing non-detection of neutrinos.

From the observational point of view, the spectrum of the majority of GRBs can be well described by the Band function \citep{1993ApJ...413..281B}, indicating a nonthermal origin. GRB 080916C is a classic example of a non-thermal dominated spectrum, which has been interpreted in the context of the internal-collision-induced magnetic reconnection and turbulence (ICMART) model with a Poynting-flux-dominated jet \citep{2009Sci...323.1688A,2009ApJ...700L..65Z,2011ApJ...726...90Z}. On the contrary, a small fraction of GRBs, such as GRB 090902B, show a spectrum with quasi-thermal component dominated, which is naturally interpreted as emission from a photosphere \citep{2009ApJ...706L.138A,2010ApJ...709L.172R}. In addition, a dominant non-thermal component and a
sub-dominant thermal component have been discovered in several GRBs, such as GRB 110721A \citep{2012ApJ...757L..31A}, GRB 160625B \citep{2017ApJ...849...71L}, and GRB 211211A \citep{2023ApJ...943..146C}. One basic question is what the difference is in the constraints on the physical parameters between the time-integrated and the summed time-resolved spectra for these three spectrally different GRBs.

The brightest-of-all-time (BOAT) GRB 221009A, on the other hand, has also been used to place significant constraints on GRB physical parameters by adopting the IceCube neutrino non-detection \citep[e.g.,][]{2022ApJ...941L..10M,2023ApJ...944..115A,2026APh...17803226P,2026ApJ..1006..201V}. However, most of these studies have adopted either time-integrated spectra \citep[e.g.,][]{2022ApJ...941L..10M,2023ApJ...944..115A,2026APh...17803226P} or relied on time-resolved spectra with Fermi/GBM data \citep[e.g.,][]{2026ApJ..1006..201V}. Due to the extreme brightness of this burst, Fermi/GBM suffered severe pulse pile-up (PPU). Although PPU reconstruction methods have been developed, they may introduce additional systematic uncertainties \citep{2026ApJ..1006..201V}. Gravitational wave high-energy Electromagnetic Counterpart All-sky Monitor (GECAM)-C, the third member of GECAM constellation, was launched onboard SATech-01 experimental satellite \citep{2023NIMPA105668586Z}. It provides prompt emission spectra without being affected by this instrumental effect \citep{2023arXiv230301203A}.

Recently, the IceCube Collaboration released a new dataset of muon track data, extending the previously published 10-year point-source dataset to cover 14 years of observations (April 6, 2008 - May 23, 2022) \citep{2026arXiv260519040A}. This release also includes updated effective areas, which provide improved sensitivity compared to the 10-year dataset \citep{2021arXiv210109836I}. In this paper, we investigate three GRBs (e.g., GRBs 080916C, 110721A, and 090902B) with different spectral components to constrain the physical parameter space by adopting the neutrino non-detection with newly released 14-year IceCube effective areas in Sec. \ref{sec:2}. In Sec. \ref{sec:3}, we use GECAM-C time-resolved spectra to calculate the neutrino emission from GRB 221009A and constrain its physical parameters. The conclusions with some discussions are presented in Sec. \ref{sec:4}. Throughout this paper, we adopt the cosmological parameters with $H_{0}=70 \text{ km s}^{-1}\text{ Mpc}^{-1}, \Omega_{m}=0.3, \Omega_{\Lambda}=0.7$, and the convention $Q_{x} = Q/10^{x}$ in centimeter gram seconds.

\section{High energy neutrinos from different spectral components of GRB}
\label{sec:2}
From the theoretical point of view, \citet{2026ApJ..1010..234Y} studied the neutrino emission from GRB jets propagating inside the cavity within the active galactic nucleus (AGN) disk by considering interactions between accelerated protons and both the prompt photons from the GRB and the radiation field of the surrounding AGN disk. Based on the calculation framework of neutrino emission from \citet{2026ApJ..1010..234Y}, we focus on studying several observed individual GRBs (e.g., GRBs 080916C, 110721A, and 090902B) with different spectral components by only considering the photon field associated with the prompt emission. The calculation for each burst is performed by ourselves using its observed spectral parameters. The detailed calculation method and the assumptions adopted are described in Appendix~\ref{app:method}. Firstly, we extract both time-resolved and time-integrated spectra for those three GRBs, and more details of spectral extraction can be found in our previous work \citep{2018ApJ...862..155L}. Then, we adopt the parameters of spectra, the observed isotropic-equivalent gamma-ray energy ($E_{\rm \gamma, iso}$) and luminosity ($L_{\rm \gamma, iso}$) to constrain their physical parameters. Finally, we compare the parameter constraints obtained from the time-resolved spectra with those obtained from the time-integrated spectra. The proton energy ($E_{p,\rm iso}$) and magnetic field luminosity ($L_{B,\rm iso}$) are related to the isotropic-equivalent gamma-ray energy and luminosity, namely, $E_{p,\rm iso} = \frac{\varepsilon_p}{\varepsilon_e} E_{\rm \gamma, iso}$, and $L_{B,\rm iso} = \frac{\varepsilon_B}{\varepsilon_e} L_{\rm \gamma, iso}$. Here, the ratios $\varepsilon_p/\varepsilon_e$ and $\varepsilon_B/\varepsilon_e$ are the fraction of jet dissipated energy going into protons and random magnetic fields relative to electrons, respectively\footnote{One needs to note that the relationship between the isotropic energy and luminosity is $E_{\rm \gamma, iso} = L_{\rm \gamma, iso} \times \Delta t / (1+z)$, where $\Delta t $ is the duration of the time interval.}. In our calculations, we assume $\varepsilon_e = \varepsilon_B$.

\subsection{GRB 080916C}
GRB 080916C is a classic case of a Poynting-flux-dominated outflow, as suggested by the lack of a thermal component in its prompt emission and its consistency with the ICMART model \citep{2011ApJ...726...90Z}. The IceCube neutrino observatory was under construction when this burst was detected, operating with a 40-string configuration. We note that the effective area of the IC40 configuration in the recently released 14-year IceCube dataset is consistent with that of the 10-year ones. Although previous studies have shown that the IC40 sensitivity is insufficient to place strong constraints on the ICMART model parameters for this burst \citep{2024ApJ...976..174O}, the IC40 effective area still provides an opportunity to investigate the impact of using time-resolved spectra on the expected neutrino event numbers across different parameter spaces.

Table \ref{table1} shows the parameters of spectral fitting for both time-resolved and time-integrated spectra. The data are obtained from Fermi Gamma-Ray Burst Monitor (GBM) and Large Area Telescope (LAT) observations, and adopt the Band function to fit \citep{1993ApJ...413..281B}. Here, $\Delta t =t_2-t_1 $, and $t_1$ and $t_2$ are the start and end times of the time interval, respectively. By assuming the radius $R_{\text{ICMART}} = 10^{16}\,\text{cm}$, Lorentz factor $\Gamma = 500$, and $\varepsilon_p/\varepsilon_e=3$, we calculate the neutrino spectra of both the time-integrated and the time-resolved spectra, as well as the summed spectrum of all time intervals, within the ICMART model in the left panel of Figure \ref{Fig:1}. We also calculate the expected number of neutrino events, about $2.46\times10^{-7}$ and $1.97\times10^{-7}$ for the summed time-resolved and the time-integrated spectra, respectively. It is found that the expected number of neutrino events from the summed time-resolved spectra is about $1.25$ times that from the time-integrated spectrum.

Moreover, we also present the model-independent constraints in the $R$-$\Gamma$ parameter space (e.g., setting free parameters for $R$ and $\Gamma$), where the color indicates the number of expected neutrino events for time-resolved spectra with $\varepsilon_p/\varepsilon_e=3$ (see middle panel of Figure \ref{Fig:1}). No matter what parameter space we select, it is found that the expected number of neutrino events is still lower than $10^{-5}$. It means that it is very difficult to detect with IceCube. The right panel of Figure \ref{Fig:1} shows the ratio between the expected neutrino event numbers calculated from the summed time-resolved spectra and those from the time-integrated spectra. We find that the expected neutrino events from the summed time-resolved spectra are slightly lower than those from the time-integrated approach in the region of relatively low $R$ and $\Gamma$, while the opposite trend is found in the high $R$ and high $\Gamma$ region. In the high-$R$ and high-$\Gamma$ region of the parameter space, the ratio is approximately 1.25, whereas in the region of relatively low $R$ and $\Gamma$ the ratio decreases to approximately 0.9.

\subsection{GRB 110721A}

The spectrum of GRB 110721A prompt emission exhibits a subdominant thermal component that is a superposition on the non-thermal Band component. This burst was detected in 2011, when IceCube was operating with the 86-string configuration. In the 10-year IceCube dataset, the IC86-2011 season (IC86\uppercase\expandafter{\romannumeral+1}) was processed with a unique procedure, resulting in an effective area that differed from that of the subsequent IC86 seasons. In the recently released 14-year dataset, however, the IC79 through IC86-2018 seasons have been reprocessed with updated calibration, resulting in a unified detector response across all IC86 seasons \citep{2026arXiv260519040A}. The updated effective area for IC86-2011 is therefore consistent with that of the other IC86 seasons, leading to a slight increase in the expected neutrino event rate compared to the 10-year dataset.

Similar to Table \ref{table1}, Table \ref{table2} shows the parameters of spectral fitting for both time-resolved and time-integrated spectra, but here the Band and blackbody models are adopted for the fits. In our calculations, both thermal and non-thermal components are assumed to originate from the same emission region, and the same values of $R_{\rm ph}$ and $\Gamma_{\rm ph}$ are applied to do the calculations for both of them. By assuming $\varepsilon_p/\varepsilon_e=3$, both $R_{\rm ph}$ and $\Gamma$ which are derived following the method in \citet{2023ApJ...943..146C} are given in Table \ref{table2}, we calculate the neutrino spectra of both time-integrated and time-resolved spectra, as well as the summed spectrum of all time intervals within the photosphere model in the left panel of Figure \ref{Fig:2}. On the other hand, we also calculate the expected number of neutrino events, about $5.30\times10^{-4}$ and $5.49\times10^{-4}$ for the summed time-resolved and the time-integrated spectra, respectively. It is found that the expected number of neutrino events from the summed time-resolved spectra is slightly lower than that from the time-integrated spectrum.

Similar to middle and right panels of Figure \ref{Fig:1}, we also present the model-independent constraints in the $R$-$\Gamma$ parameter space (see middle panel of Figure \ref{Fig:2}), and the ratio between the expected neutrino event numbers which are calculated from the summed time-resolved spectra and those from the time-integrated spectra (see right panel of Figure \ref{Fig:2}). It is found that no matter what parameter space we select, the expected number of neutrino events is still lower than $10^{-3}$, and it is still difficult to detect with IceCube. Moreover, the expected neutrino events from the summed time-resolved spectra are still slightly lower than those from the time-integrated approach in the region of relatively low $R$ and $\Gamma$. In the high-$R$ and high-$\Gamma$ region of the parameter space, the ratio is approximately 1.4, whereas in the region of relatively low $R$ and $\Gamma$ the ratio decreases to approximately 0.7.

\subsection{GRB 090902B}
GRB 090902B has a thermal-dominated spectrum, making it a representative of the photosphere-origin scenario. The prompt emission spectrum is well described by a blackbody together with a power-law model \citep{2009ApJ...706L.138A,2010ApJ...709L.172R, 2011ApJ...730..141Z}. The IceCube neutrino observatory was operating with the 59-string configuration (IC59) when this burst was detected. We note that the effective area of the IC59 configuration in the recently released 14-year IceCube dataset is consistent with that of the 10-year dataset.

Table \ref{table3} shows the parameters of spectral fitting for both time-resolved and time-integrated spectra by adopting the blackbody and power-law models to fit. By assuming $\varepsilon_p/\varepsilon_e=3$ and derived $R_{\rm ph}$ and $\Gamma$ given in Table \ref{table3}, the neutrino spectra of both time-integrated and time-resolved spectra, as well as the summed spectrum of all time intervals within the photosphere model are shown in the left panel of Figure \ref{Fig:3}. The expected number of calculated neutrino events is about $1.31\times10^{-3}$ and $1.01\times10^{-3}$ for the summed time-resolved and the time-integrated spectra, respectively. It is found that the expected number of neutrino events from the summed time-resolved spectra is about 1.3 times that from the time-integrated spectrum.

The model-independent constraints in the $R$-$\Gamma$ parameter space are also shown in the middle panel of Figure \ref{Fig:3}. The expected number of neutrino events is still lower than $3\times10^{-2}$, and it is difficult to detect with IceCube. Similar to the right panel of Figure \ref{Fig:1}, it is also found that the expected neutrino events from the summed time-resolved spectra are slightly lower than those from the time-integrated approach in the region of relatively low $R$ and $\Gamma$. In the high-$R$ and high-$\Gamma$ region of the parameter space, the ratio is approximately 2.1, whereas in the region of relatively low $R$ and $\Gamma$ the ratio decreases to approximately 0.9. On the other hand, we also note that the derived Lorentz factor exceeds 1000 in some bright time intervals. Such large Lorentz factors place GRB 090902B in the high-$\Gamma$ and relatively small-$R$ part of the parameter space, where the ratio between the summed time-resolved and time-integrated results is approximately 1.3.

\section{Constraint on the parameter space of GRB 221009A}
\label{sec:3}
In this section, we calculate the neutrino emission and derive constraints on the physical parameters of GRB 221009A by using the GECAM-C data and the newly released 14-year IceCube effective area. We divide the prompt emission into 43 time intervals and find that the Band function describes both the time-resolved and the time-integrated spectra of the GRB 221009A prompt emission well. Based on the spectral parameters obtained from fits, one can roughly calculate both $E_{\rm \gamma, iso}$ and $L_{\rm \gamma, iso}$ for each time interval. More details can be found in \cite{Zhang2026}. Unsaturated measurements of GECAM-C data can provide more reliable constraints on the physical parameters than those obtained in previous studies, which relied on Fermi/GBM data affected by pulse pile-up.

In order to constrain the physical parameters of GRB 221009A, we consider two scenarios of its prompt emission. One is to adopt a model-independent framework and place constraints in the $R$-$\Gamma$ parameter space. The other one is to consider the dissipation radius which is related to the bulk Lorentz factor $\Gamma$ and the minimum variability timescale $\delta t_{\rm min}$ by $R_{\rm IS} = 2\Gamma^2 c \delta t_{\rm min} / (1+z)$ within the internal shock model.

The left panel of Figure~\ref{Fig:4} also shows the time-resolved neutrino spectra under the ICMART model, with $R_{\text{ICMART}} = 10^{16}\,\text{cm}$, $\Gamma = 500$ and $\varepsilon_p/\varepsilon_e=3$. The expected number of neutrino events is $1.02\times10^{-1}$ for the summed time-resolved spectra and $3.36\times10^{-2}$ for the time-integrated spectra. The model-independent constraints in the $R$-$\Gamma$ parameter space are shown in the middle panel of Figure~\ref{Fig:4} for given different values of $\varepsilon_p/\varepsilon_e$. The right panel of Figure~\ref{Fig:4} shows the ratio between the expected neutrino event numbers from the summed time-resolved spectra and those from the time-integrated spectra. For example, the constraints from the time-resolved spectra are stronger than those from the time-integrated analysis at $R = 1.07\times10^{15}\,\text{cm}$. For given increasing value of $\varepsilon_p/\varepsilon_e = 1, 3$ and $10$, the constrained lower limits of the Lorentz factor are also increased as $\Gamma > 254$, $\Gamma > 371$, and $\Gamma > 537$ of time-resolved spectra, and $\Gamma > 188$, $\Gamma > 273$, and $\Gamma > 395$ of time-integrated spectra.



It is found that the parameter space of the ICMART model can be constrained by the neutrino non-detection of GRB 221009A. The value of the ratio varies across the parameter space, ranging from about 0.6 at low $R$ and low $\Gamma$ to nearly 3 at high $R$ and high $\Gamma$. The time-resolved spectra generally provide stronger constraints than the time-integrated spectra when the dissipation radius is relatively large. However, for $\varepsilon_p/\varepsilon_e = 1$, with low $R$ and low $\Gamma$, the time-integrated spectra actually exclude a larger parameter space than the time-resolved ones. For relatively high values of $\varepsilon_p/\varepsilon_e$ (e.g., $\varepsilon_p/\varepsilon_e = 10$), the time-resolved spectra produce lower neutrino event numbers than those of the time-integrated ones, but these values still exceed the $N_{\rm sig}=2.3$ threshold, so the corresponding parameter regions can still be excluded.

On the other hand, the prompt emission of GRB 221009A may originate from internal shocks. Figure~\ref{Fig:5} shows the constraints in the $\Gamma$-$\varepsilon_p/\varepsilon_e$ parameter space (left panel) and the ratio of the expected neutrino event numbers from the time-resolved spectrum to those from the time-integrated spectrum (right panel) within the internal shock model. It is found that the ratio only depends on the value of $\Gamma$, and does not depend on the value of $\varepsilon_p/\varepsilon_e$. Within the internal shock framework for $\Gamma = 400$ and 700, one can obtain $\varepsilon_p/\varepsilon_e < 0.66$ and $\varepsilon_p/\varepsilon_e < 6.98$ for time-resolved spectra, and $\varepsilon_p/\varepsilon_e < 1.30$ and $\varepsilon_p/\varepsilon_e < 19.20$ for time-integrated spectra, respectively. The expected neutrino events from time-resolved spectra exceed those from time-integrated spectra by a factor of about three in the high-$\Gamma$ and high-$R$ regime.


\section{Conclusion and discussion}\label{sec:4}
The non-detection of neutrinos associated with GRBs has been used to constrain the physical parameter space of GRB jet models. In this paper, by selecting three peculiar GRBs whose spectra of prompt emission are composed of different components, such as GRB 080916C (non-thermal emission dominated), GRB 110721A (thermal + non-thermal emission), and GRB 090902B (thermal emission dominated), we investigate the impact of time-resolved spectra of prompt emission for those three GRBs on calculated neutrino emission and physical parameter constraints. Moreover, the BOAT GRB 221009A using the GECAM-C data which provide undistorted measurements of the prompt emission, is also used to calculate the expected neutrino events.

By analyzing both time-resolved and time-integrated spectra of prompt emission for those GRBs, we calculate the expected neutrino events and constrain their physical parameters. Several numerical calculations are summarized as follows:
\begin{itemize}
 \item The expected neutrino events from time-resolved spectra of prompt emission for GRBs 090902B, 110721A and 080916C are lower than those from time-integrated spectra at low $R$ (corresponding to the photosphere model ones) and low $\Gamma$, while the opposite is true at high $R$ (corresponding to the ICMART model ones) and high $\Gamma$.
 \item For GRB 221009A, the model-independent constraints set lower limits on the bulk Lorentz factor for different $R$, excluding a large parameter space (including the low $\Gamma$ region relevant for the ICMART model). In general, the time-resolved spectra provide stronger constraints than those of the time-integrated ones. However, in the region of low $R$ and low $\Gamma$ with a low $\varepsilon _{\rm p}/\varepsilon_{\rm e}$, the time-integrated spectra exclude a larger parameter space.
 \item Within the internal shock model, we place upper limits on $\varepsilon_p/\varepsilon_e$ for different $\Gamma$. The ratio of expected neutrino event numbers depends only on $\Gamma$ and does not depend on $\varepsilon_p/\varepsilon_e$.
\end{itemize}

We also checked how strongly the results depend on the assumed proton spectral index. Using the same calculation method, we compute the expected number of events for GRB 080916C at $R = 10^{16}\,\text{cm}$ and $\Gamma = 500$ for $s_p = 1$, $1.5$, $2$, $2.5$ and $3$, and obtain $9.30\times10^{-9}$, $8.50\times10^{-8}$, $2.46\times10^{-7}$, $7.05\times10^{-9}$ and $6.33\times10^{-11}$, respectively. The absolute numbers therefore change by about four orders of magnitude, while the ratio between the time-resolved and time-integrated results changes only from $1.23$ to $1.27$ and remains between $0.89$ and $1.32$ over the whole parameter space.

The trend of ratio in model-independent constraints in the $R$-$\Gamma$ parameter space may result from a competition between two effects. (1) In the region of low $R$ and low $\Gamma$, although the $p\gamma$ efficiency may already be close to unity, the magnetic field inferred from the time-resolved spectrum is stronger than that from the time-integrated spectrum, leading to efficient synchrotron cooling of charged pions and muons, which may suppress the neutrino emission before its decay. (2) In the region of high $R$ and high $\Gamma$, synchrotron cooling is not significant, and the increased photon density captured by the time-resolved spectrum can effectively enhance the $p\gamma$ efficiency, leading to more efficient neutrino emission.

In order to make this difference quantitative, we compare the lower limits on the bulk Lorentz factor of those two analyses for fixed dissipation radius. For given $R \simeq 1.07\times10^{15}\,\text{cm}$ of GRB 221009A, the model-independent constraints from the time-resolved spectra yield $\Gamma > 254$, $\Gamma > 371$ and $\Gamma > 537$ for $\varepsilon_p/\varepsilon_e = 1$, $3$ and $10$, respectively. For the time-integrated spectra, the corresponding limits are $\Gamma > 188$, $\Gamma > 273$ and $\Gamma > 395$. Within the whole parameter space, the ratio between the two limits of bulk Lorentz factor ranges from $1.28$ to $1.40$. Therefore, the constraints of lower limits of bulk Lorentz factor from the time-resolved are $28\%$--$40\%$ higher than that of time-integrated analysis.

Within the internal shock model, the constrained quantity, in fact, is given an upper limit on $\varepsilon_p/\varepsilon_e$. For fixed bulk Lorentz factor, the time-resolved spectra always give a tighter upper limit. By defining the ratio between the value of $\varepsilon_p/\varepsilon_e$ from time-resolved spectra and the value of $\varepsilon_p/\varepsilon_e$ from time-integrated spectra, it is found that the ratio decreases from $0.86$ (for $\Gamma \simeq 160$) to $0.33$ (for $\Gamma = 1000$). Therefore, the two methods (e.g., time-resolved and time-integrated) differ by a factor of 1.2 to 3. For a given minimum variability timescale of $0.01\,\text{s}$, we obtain $\varepsilon_p/\varepsilon_e < 0.66$ from the time-resolved spectra and $\varepsilon_p/\varepsilon_e < 1.30$ from the time-integrated spectra for $\Gamma = 400$, and $\varepsilon_p/\varepsilon_e < 6.98$ and $< 19.20$ for $\Gamma = 700$. Since a smaller upper limit corresponds to a stronger constraint, the time-resolved analysis therefore tightens the constraint by a factor of two to three in most of the parameter space, while near $\Gamma\simeq160$ the two limits differ by only about $15\%$.


On the other hand, although we have used the newly released 14-year IceCube effective area and GECAM-C data of GRB 221009A, the model-independent constraints obtained in this work are not significantly stronger than those derived from time-resolved Fermi/GBM data in previous studies \citep{2026ApJ..1006..201V}. There are several reasons that may lead to this result: (1) The treatment of the magnetic field energy fraction is different between the two studies. A fixed value of $\varepsilon_B/\varepsilon_e = 0.01$ was adopted in \citet{2026ApJ..1006..201V}, and they claim that the value of $\varepsilon_B/\varepsilon_e$ only slightly affects the calculated number of neutrinos. We adopt the varying values of $\varepsilon_B/\varepsilon_e$ (e.g., from 0.01 to 1) to do the calculations, and we find that the effect on the neutrino event number is indeed not strong enough at high-$R$ and high-$\Gamma$ regime, but significantly affects the results at low-$R$ and low-$\Gamma$ regime. (2) the Fermi/GBM data used in previous studies suffered from severe pulse pile-up due to the extreme brightness of GRB 221009A. Although correction methods were applied, they may introduce systematic uncertainties. In contrast, the GECAM-C data used in this work provide undistorted measurements of the prompt emission, which may reduce the corresponding systematic uncertainties. (3) There is a difference in the physical processes considered to do the neutrino production calculations. In \citet{2026ApJ..1006..201V}, synchrotron cooling of secondary particles (pions and muons) was included, but proton synchrotron cooling was not taken into account. In our calculations, we include proton synchrotron cooling, which may reduce the maximum proton energy and thus affect the neutrino production.


In summary, our study demonstrates that time-resolved spectra provide a more realistic estimate of neutrino emission from GRBs, and can significantly affect the derived parameter constraints. We strongly suggest that one can adopt time-resolved spectra to do the calculations of GRB neutrino studies in the future when the data quality is good enough.

\begin{acknowledgments}
This work is supported by the Natural Science Foundation of Guangxi (grant No. 2025GXNSFDA02850010), the Natural Science Foundation of China (grant Nos. 12494574 and 12621007), the Program of Bagui Scholars Program (LHJ), the Guangxi Talent Program (Highland of Innovation Talents), and Guangxi Science and Technology Innovation Platform Program (Leitai Action Plan) (Grant No. Guike LT2600640026 ).

\end{acknowledgments}

\appendix
\section{The method of neutrino calculations }\label{app:method}

In this appendix, we present the details of neutrino calculations we used. One stipulates that the primed and unprimed quantities refer to the jet comoving frame and the burst rest frame, respectively. Quantities measured at the Earth are marked with the subscript ``obs''. These frames are related by $\varepsilon'_\gamma=\varepsilon_\gamma/\Gamma$ and $\varepsilon_\gamma=(1+z)\varepsilon_{\gamma,\rm obs}$.

We adopt the calculation framework of \citet{2026ApJ..1010..234Y}. The neutrinos are assumed to be produced through photomeson ($p\gamma$) interactions in the prompt emission region, located at radius $R$, where the jet has a bulk Lorentz factor $\Gamma$. Since the bursts studied here are not associated with AGN disks, the target photons are those of the prompt emission itself, so that no additional photon field needs to be invoked. Each time interval is treated as an independent emission episode, and the neutrino emission of that interval is computed only from its own spectral parameters. Proton-proton ($pp$) interactions are neglected, and the secondary pions and muons are assumed to lose energy only through synchrotron radiation and adiabatic expansion.

The spectral shape of the target photons is taken directly from the spectral fit, i.e., a Band function \citep{1993ApJ...413..281B} for GRBs 080916C and 221009A, a Band function plus a blackbody for GRB 110721A, and a blackbody plus a power law for GRB 090902B. The normalization is fixed by the comoving photon energy density,
\begin{equation}
U'_\gamma=\frac{L_{\rm\gamma,iso}}{4\pi c\Gamma^2R^2}
=\int_{\varepsilon'_{\gamma,m}}^{\varepsilon'_{\gamma,M}}
\varepsilon'_\gamma\,n'_\gamma(\varepsilon'_\gamma)\,d\varepsilon'_\gamma,
\label{eq:Ugamma}
\end{equation}
with $\varepsilon'_{\gamma,m}=0.1$ eV and $\varepsilon'_{\gamma,M}=1$ MeV \citep{2006PhRvD..73f3002M}.

The magnetic field strength is obtained from $L_{\rm\gamma,iso}$ and the assumed ratio of magnetic to electron energy,
\begin{equation}
B'=\left[\frac{2(\varepsilon_B/\varepsilon_e)L_{\rm\gamma,iso}}{c\Gamma^2R^2}\right]^{1/2}.
\label{eq:Bfield}
\end{equation}

Protons are injected with a power-law distribution and an exponential cutoff,
\begin{equation}
\frac{dN_p}{dE_p}=A_p\,E_p^{-s_p}\exp\left(-\frac{E_p}{E_{p,\max}}\right),
\label{eq:proton}
\end{equation}
where $A_p$ is the normalization factor of the proton spectrum, $E_{p,\rm iso}=(\varepsilon_p/\varepsilon_e)E_{\rm\gamma,iso}$ is the total proton energy, and $s_p$ is the proton spectral index, for which we adopt $s_p=2$. The minimum proton energy is set by $E'_{p,\min}=10\,m_pc^2$, i.e., $E_{p,\min}=\Gamma E'_{p,\min}$, while the maximum proton energy $E_{p,\max}$ is determined below.

The photomeson production rate at a given proton energy is
\begin{equation}
t_{p\gamma}^{\prime-1}=\frac{c}{2\gamma_p^{\prime2}}
\int_{\bar{\varepsilon}_{\rm th}}^{\infty}d\bar{\varepsilon}_\gamma\,
\sigma_{p\gamma}\kappa_{p\gamma}\bar{\varepsilon}_\gamma
\int_{\bar{\varepsilon}_\gamma/2\gamma'_p}^{\infty}
d\varepsilon'_\gamma\,\varepsilon_\gamma^{\prime-2}n'_\gamma(\varepsilon'_\gamma),
\label{eq:tpgamma}
\end{equation}
where $\gamma_p'=E'_p/(m_pc^2)$ is the Lorentz factor of the proton, $\bar{\varepsilon}_\gamma$ is the photon energy in the proton rest frame, and $\bar{\varepsilon}_{\rm th}=145$ MeV is the threshold energy of photomeson production. The cross section $\sigma_{p\gamma}$ is taken from the fitting formulae of \citet{Kossov(2002)}, which are based on GEANT4, and the inelasticity is approximated by the piecewise function $\kappa_{p\gamma}=0.2$ for $\bar{\varepsilon}_\gamma<983$ MeV and $\kappa_{p\gamma}=0.6$ for $\bar{\varepsilon}_\gamma\geq983$ MeV \citep{Atoyan-Dermer(2001)}.
The maximum proton energy is then obtained by equating the acceleration and cooling timescales,
\begin{equation}
t'_{p,\rm acc}=t'_{p,\rm cool},\qquad
t'_{p,\rm acc}=\frac{E'_p}{ecB'},\qquad
t'^{-1}_{p,\rm cool}=t'^{-1}_{\rm dyn}+t'^{-1}_{p,\rm syn}+t'^{-1}_{p\gamma},
\label{eq:acc_eq_cool}
\end{equation}
where $E'_p=E_p/\Gamma$ is the comoving proton energy, $t'_{\rm dyn}=R/(\Gamma c)$, and $t'_{p,\rm syn}=6\pi m_p^4c^3/(m_e^2\sigma_{\rm T}B'^2E'_p)$.

The fraction of the proton energy that is converted into pions, and the suppression factors caused by the synchrotron cooling of pions and muons before their decay, are
\begin{equation}
f_{p\gamma}=\frac{t'_{p,\rm cool}}{t'_{p\gamma}},\qquad
f_{\rm sup,\pi}=1-\exp\left(-\frac{t'_{\pi,\rm cool}}{t'_{\pi,\rm dec}}\right),\qquad
f_{\rm sup,\mu}=1-\exp\left(-\frac{t'_{\mu,\rm cool}}{t'_{\mu,\rm dec}}\right),
\label{eq:fsup}
\end{equation}
where $E'_\pi=E'_p/5$ and $E'_\mu=0.15E'_p$ are the pion and muon energies, with Lorentz factors $\gamma'_\pi=E'_\pi/(m_\pi c^2)$ and $\gamma'_\mu=E'_\mu/(m_\mu c^2)$, and decay timescales $t'_{\pi,\rm dec}=\gamma'_\pi\tau_\pi$ and $t'_{\mu,\rm dec}=\gamma'_\mu\tau_\mu$. Here $\tau_\pi=2.6\times10^{-8}$~s and $\tau_\mu=2.2\times10^{-6}$~s. The corresponding cooling timescales are $t'^{-1}_{\pi,\rm cool}=t'^{-1}_{\pi,\rm syn}+t'^{-1}_{\rm dyn}$ and $t'^{-1}_{\mu,\rm cool}=t'^{-1}_{\mu,\rm syn}+t'^{-1}_{\rm dyn}$, with the synchrotron timescale given by the same expression as $t'_{p,\rm syn}$ for the corresponding mass and energy. The definitions of the suppression factors follow \citet{2006PhRvD..73f3002M}.

The neutrino spectra from pion and muon decay, with $E_\nu\simeq0.05E_p$, are
\begin{equation}
E_\nu^2\frac{dN_{\nu_\mu}}{dE_\nu}=\frac{1}{8}f_{p\gamma}f_{\rm sup,\pi}\,E_p^2\frac{dN_p}{dE_p},
\qquad
E_\nu^2\frac{dN_{\bar{\nu}_\mu}}{dE_\nu}=E_\nu^2\frac{dN_{\nu_e}}{dE_\nu}
=\frac{1}{8}f_{p\gamma}f_{\rm sup,\pi}f_{\rm sup,\mu}\,E_p^2\frac{dN_p}{dE_p},
\label{eq:numu}
\end{equation}
These expressions follow \citet{1997PhRvL..78.2292W} and \citet{2006PhRvD..73f3002M}.

Before neutrino mixing, the quantity $E_\nu^2\Phi^0_{\nu_i}$ in the burst rest frame is
\begin{equation}
E_\nu^2\Phi^0_{\nu_i}=\frac{1}{4\pi d_L^2}E_\nu^2\frac{dN_{\nu_i}}{dE_\nu},
\label{eq:phi0}
\end{equation}
where $d_L$ is the luminosity distance and the superscript $0$ denotes quantities before neutrino mixing.
After oscillation during propagation, the quantity $E_\nu^2\Phi_{\nu_\mu}$ in the burst rest frame is
\begin{equation}
E_\nu^2\Phi_{\nu_\mu}=\frac{4}{18}E_\nu^2\Phi^0_{\nu_e+\bar{\nu}_e}
+\frac{7}{18}\left(E_\nu^2\Phi^0_{\nu_\mu+\bar{\nu}_\mu}+E_\nu^2\Phi^0_{\nu_\tau+\bar{\nu}_\tau}\right),
\label{eq:phi}
\end{equation}
The coefficients correspond to tri-bimaximal mixing \citep{Harrison-et-al.(2002)}. Since tau neutrinos are not produced at the source in this approximation, $\Phi^0_{\nu_\tau}=\Phi^0_{\bar{\nu}_\tau}=0$. Finally, the neutrino energies and the fluence are converted to the values measured at the Earth through $E_{\nu,\rm obs}=E_\nu/(1+z)$ and $\Phi_{\rm obs}=(1+z)\Phi$, so that the quantity shown in the figures is $(E^2\Phi)_{\rm obs}=(E^2\Phi)/(1+z)$.

The expected number of muon-neutrino events is obtained by convolving the fluence with the effective area of the detector,
\begin{equation}
N_{\rm sig}=\int\Phi_{\nu_\mu}(E_{\nu,\rm obs})\,A_{\rm eff}(E_{\nu,\rm obs},\delta)\,dE_{\nu,\rm obs},
\label{eq:Nsig}
\end{equation}
where $A_{\rm eff}$ is taken from the 14-year IceCube point-source dataset \citep{2026arXiv260519040A} and $\delta$ is the declination of the source.

\begin{figure*}
      \centering
      
      \includegraphics[width=0.34\linewidth]{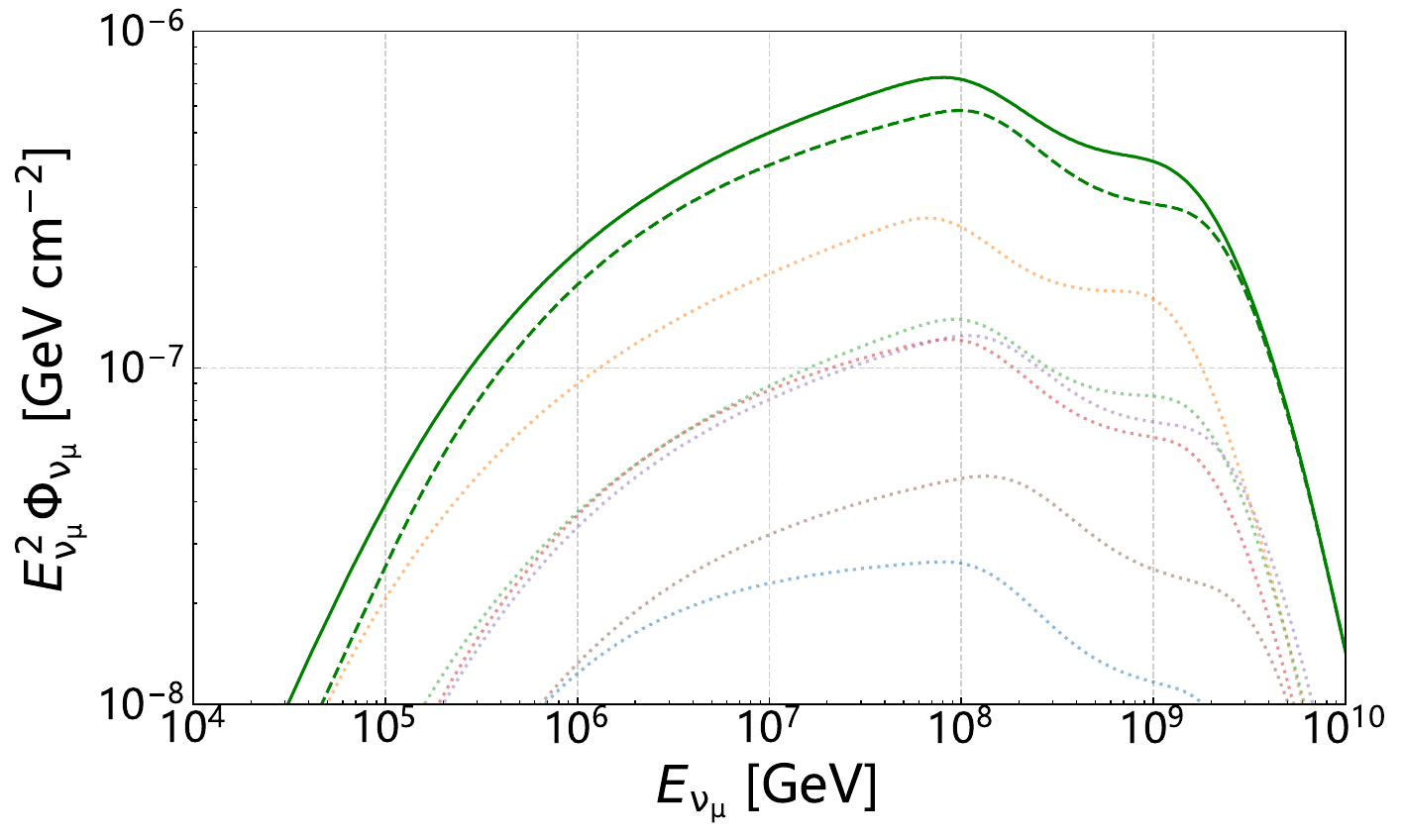}
      \includegraphics[width=0.32\linewidth]{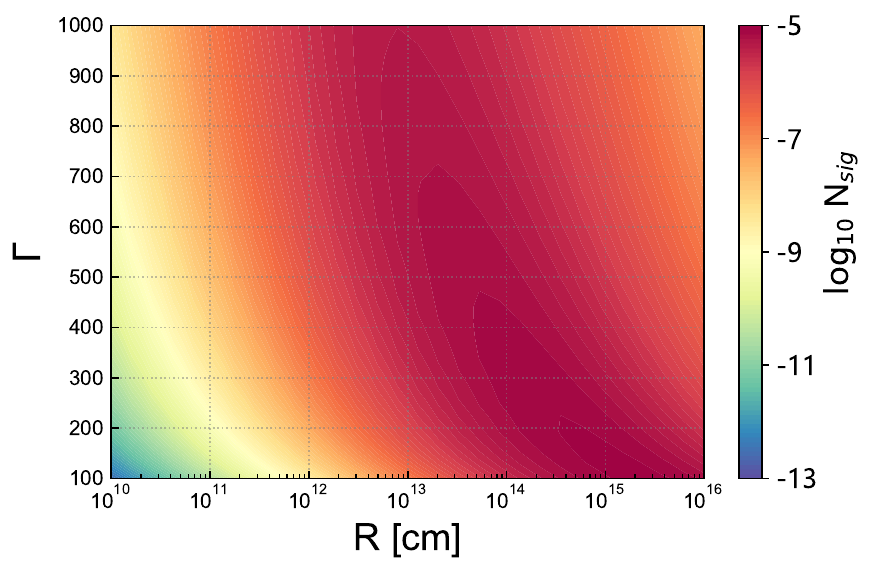}
      \includegraphics[width=0.32\linewidth]{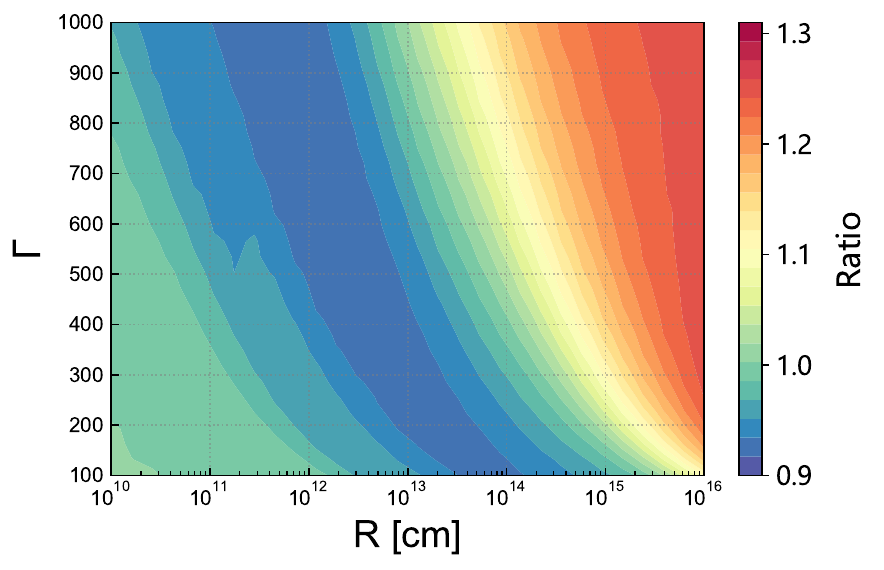}

    \caption{Left: neutrino spectra as a function of the observed neutrino energy of the time-integrated spectrum (dashed line), of the individual time intervals (dotted lines of different colors, one for each interval), and their sum (solid line) in the ICMART model for GRB 080916C with $\varepsilon_p/\varepsilon_e=3$, $R_{\text{ICMART}} = 10^{16}\,\text{cm}$ and $\Gamma = 500$. The neutrino energy shown is the observed one, and the vertical axis is $E_{\nu_\mu}^2\Phi_{\nu_\mu}$ in units of GeV cm$^{-2}$. Middle: model-independent constraints in the $R$-$\Gamma$ parameter space, where the color indicates the expected number of neutrino events $N_{\rm sig}$ obtained from the time-resolved spectra. Right: ratio between the expected neutrino event numbers from the summed time-resolved spectra and those from the time-integrated spectra, where the color indicates the value of this ratio.}
      \label{Fig:1}
  \end{figure*}

\begin{figure*}
      \centering
      
      \includegraphics[width=0.32\linewidth]{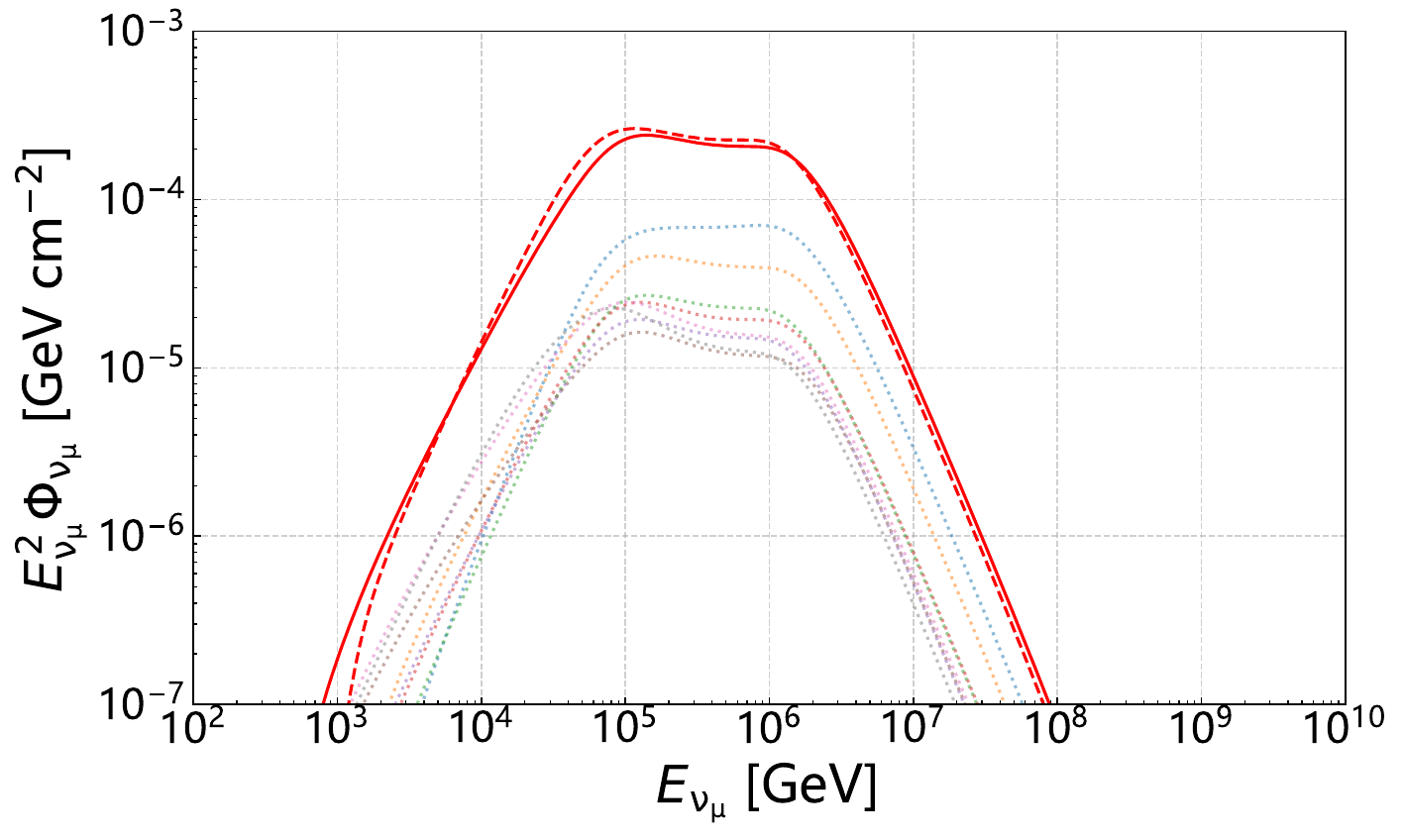}
      \includegraphics[width=0.32\linewidth]{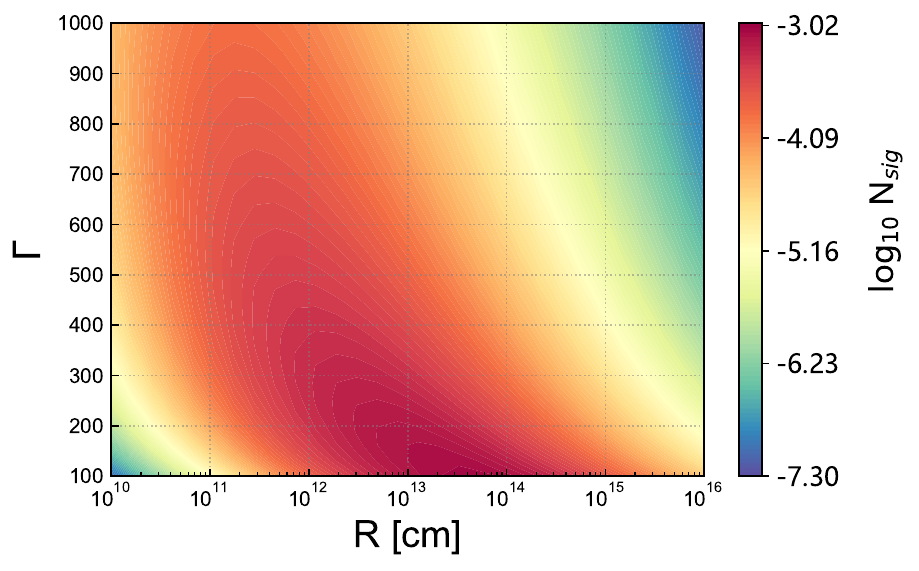}
      \includegraphics[width=0.32\linewidth]{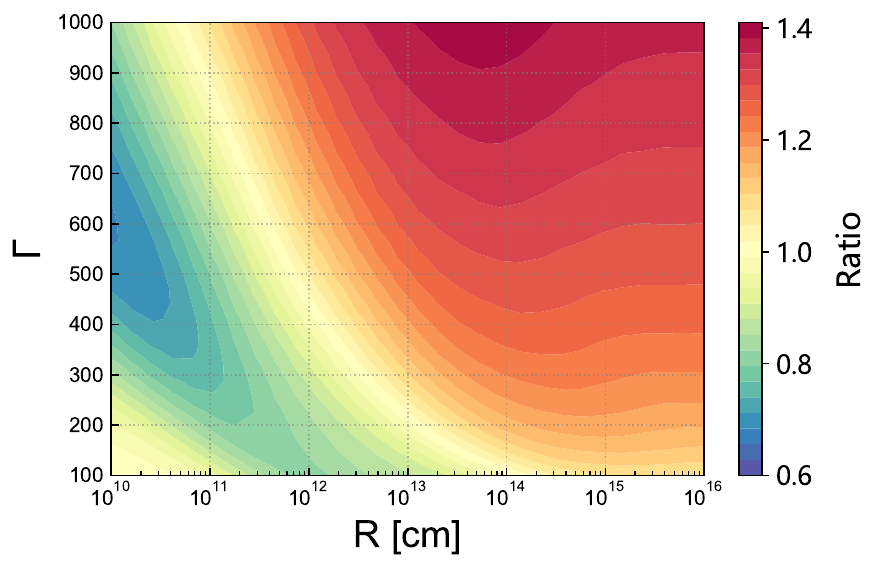}

      \caption{Similar to Figure \ref{Fig:1}, but for GRB 110721A within the photosphere model and given values of $R_{\rm ph}$ and $\Gamma_{\rm ph}$ in Table \ref{table2}. The meaning of the lines, of the colors, and of the axes is the same as in Figure \ref{Fig:1}.}
      \label{Fig:2}
  \end{figure*}

\begin{figure*}
      \centering
      
      \includegraphics[width=0.32\linewidth]{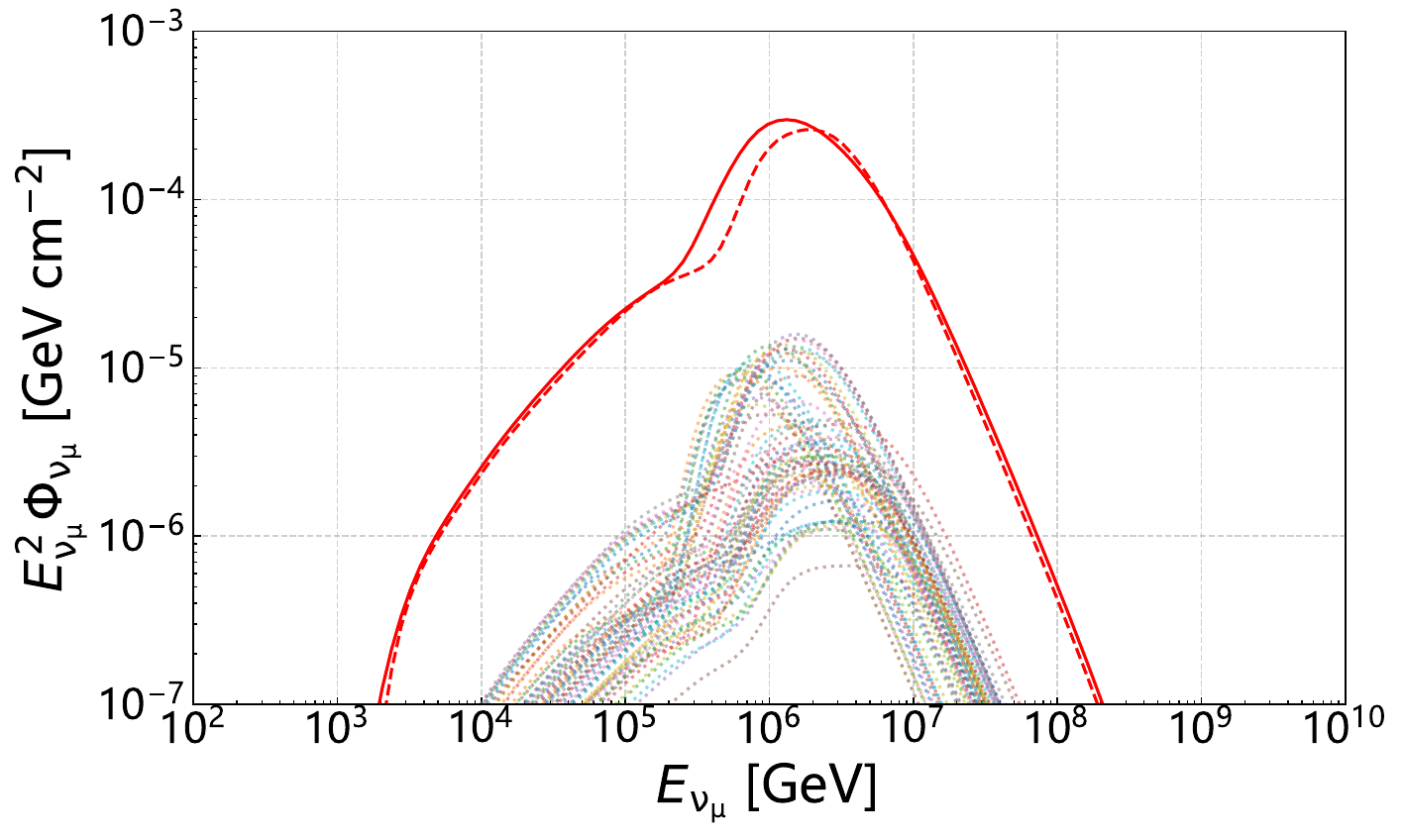}
      \includegraphics[width=0.32\linewidth]{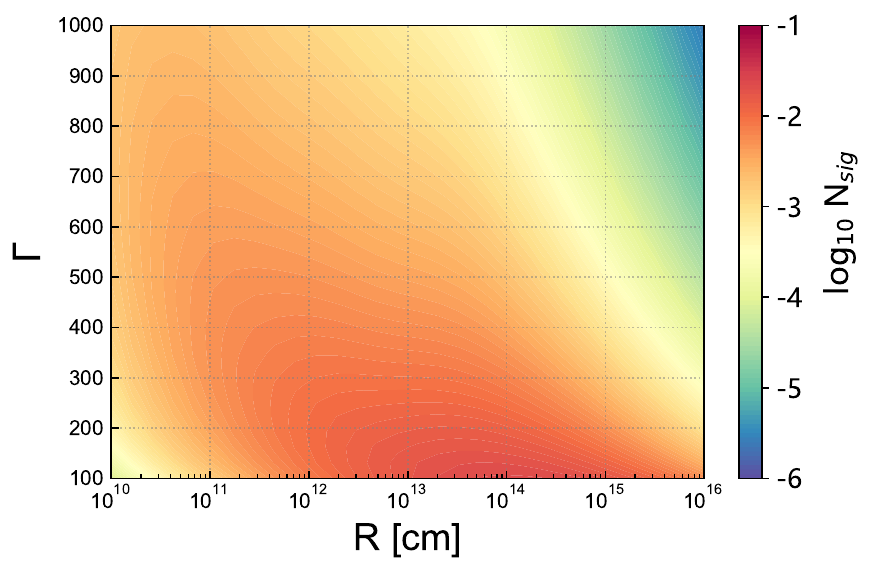}
      \includegraphics[width=0.32\linewidth]{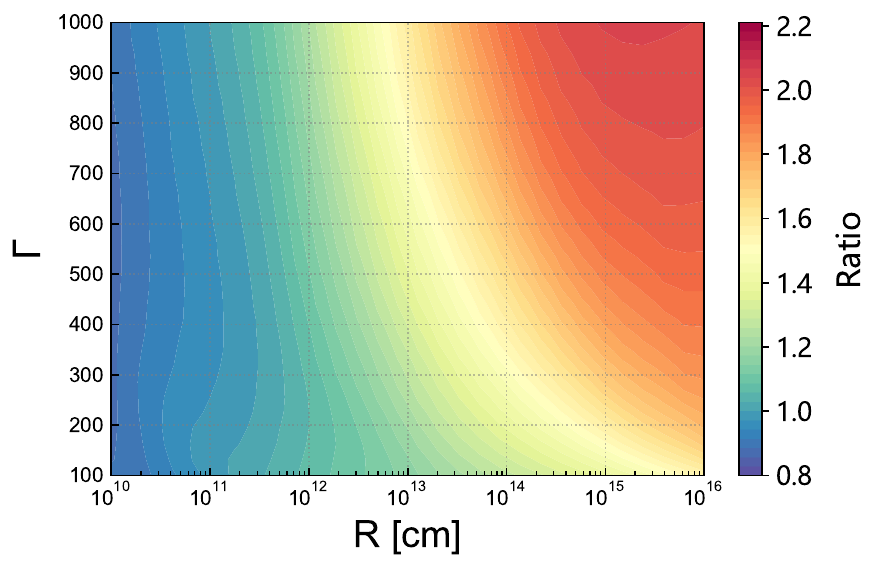}

      \caption{Similar to Figure \ref{Fig:2}, but for GRB 090902B with given values of $R_{\rm ph}$ and $\Gamma_{\rm ph}$ in Table \ref{table3}. The meaning of the lines, of the colors, and of the axes is the same as in Figure \ref{Fig:1}.}
      \label{Fig:3}
  \end{figure*}

\begin{figure*}
      \centering
      
      \includegraphics[width=0.32\linewidth]{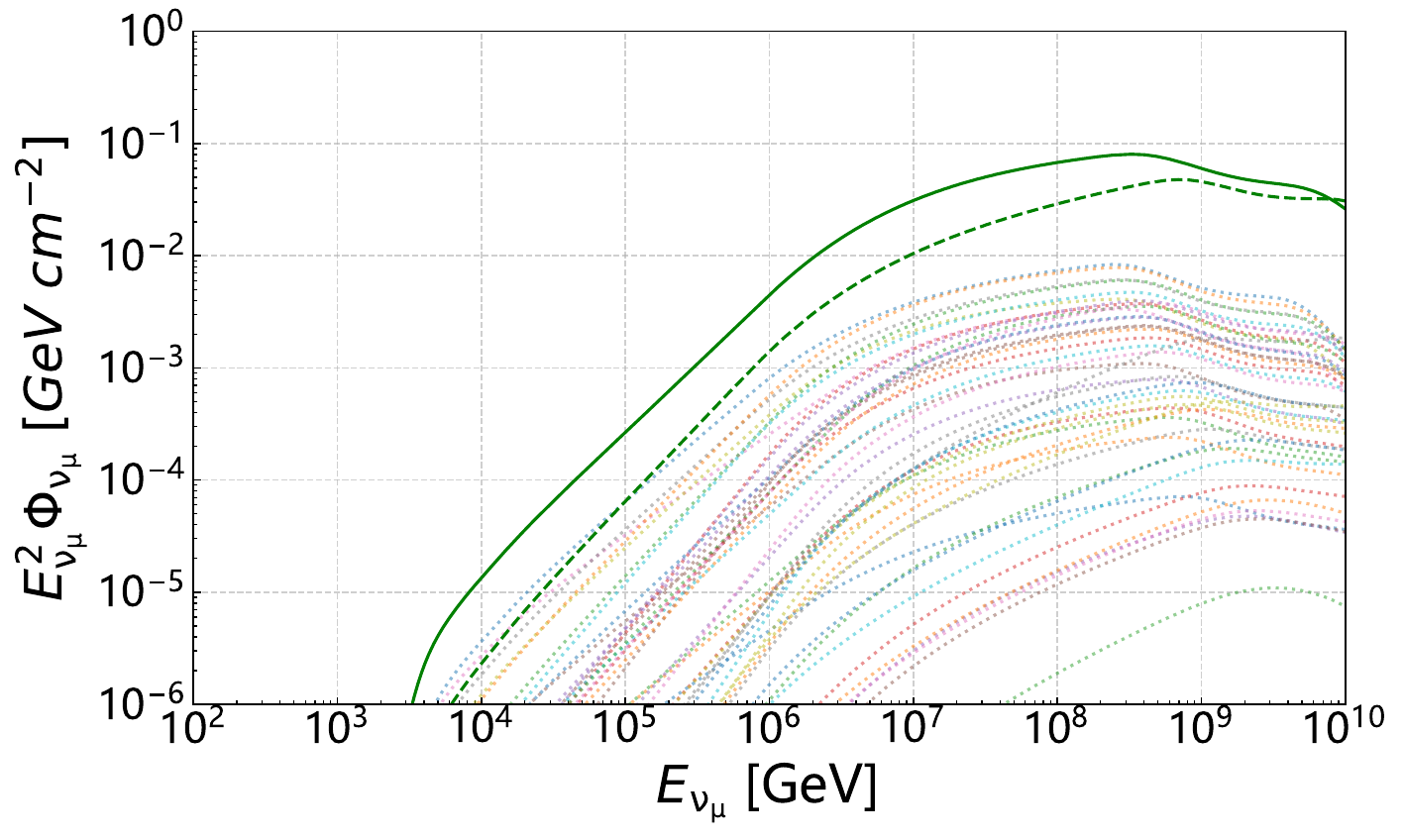}
      \includegraphics[width=0.32\linewidth]{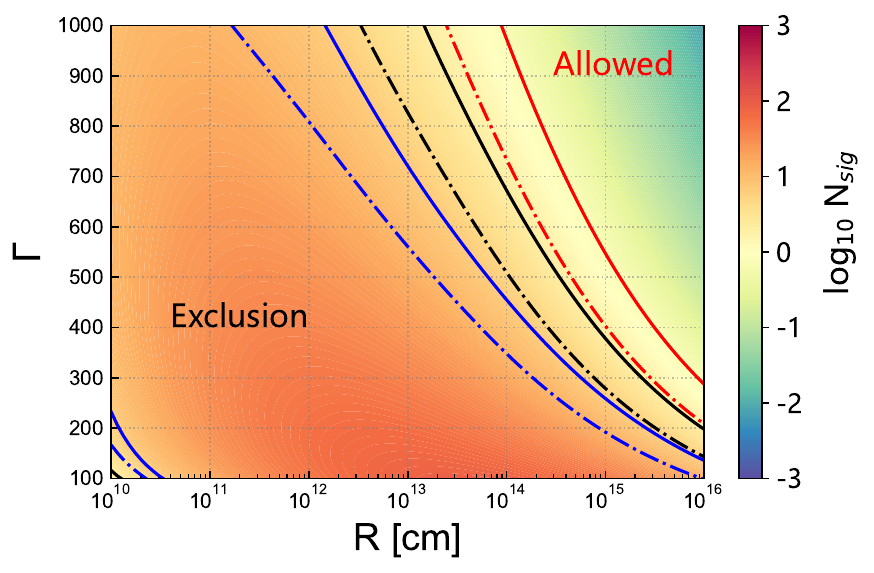}
      \includegraphics[width=0.32\linewidth]{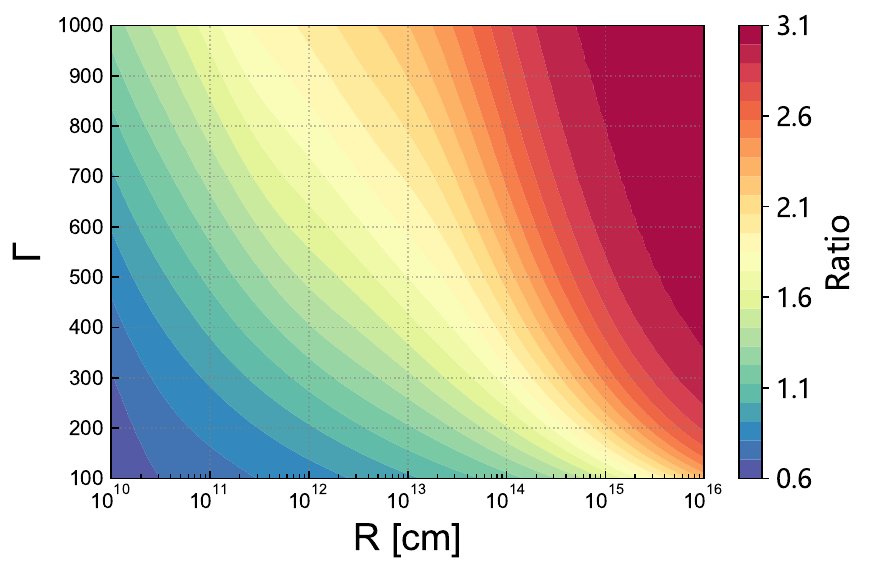}

      \caption{Left: neutrino spectra as a function of the observed neutrino energy of the time-integrated spectrum (dashed line), of the individual time intervals (dotted lines of different colors, one for each interval), and their sum (solid line) in the ICMART model for GRB 221009A with $\varepsilon_p/\varepsilon_e=3$, $R_{\text{ICMART}} = 10^{16}\,\text{cm}$ and $\Gamma = 500$. The neutrino energy shown is the observed one, and the vertical axis is $E_{\nu_\mu}^2\Phi_{\nu_\mu}$ in units of GeV cm$^{-2}$. Middle: model-independent constraints in the $R$-$\Gamma$ parameter space. The color scale represents the expected number of neutrino events for $\varepsilon_p/\varepsilon_e = 3$. The solid and dash-dotted lines represent the $N_{\rm sig}=2.3$ contours (90\% C.L.) obtained from time-resolved and time-integrated spectra, respectively, with $\varepsilon_p/\varepsilon_e = 1$ (blue), $3$ (black), and $10$ (red). Right: ratio between the expected neutrino event numbers from the summed time-resolved spectra and those from the time-integrated spectra, where the color indicates the value of this ratio.}
      \label{Fig:4}
  \end{figure*}

\begin{figure*}
      \centering
      \includegraphics[width=0.45\linewidth]{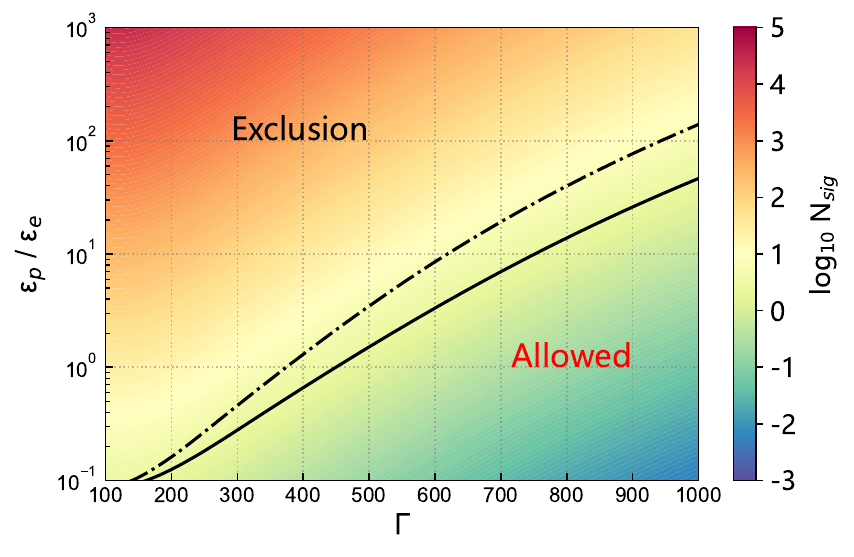}
      \includegraphics[width=0.45\linewidth]{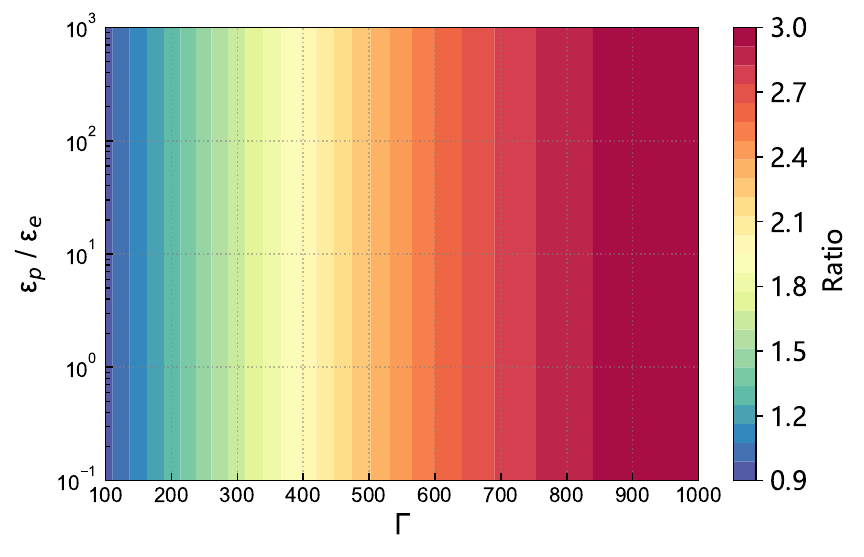}
      \caption{Similar to Figure \ref{Fig:4} for GRB 221009A, but within the internal shock model. Left: constraints in the $\Gamma$-$\varepsilon_p/\varepsilon_e$ parameter space. The solid and dash-dotted lines represent the $N_{\rm sig}=2.3$ contours (90\% C.L.) obtained from time-resolved and time-integrated spectra, respectively, with a minimum variability timescale of $\delta t_{\rm min} = 0.01\,\text{s}$. Right: ratio between the expected neutrino event numbers from the summed time-resolved spectra and those from the time-integrated spectra.}
      \label{Fig:5}
  \end{figure*}

For the time-resolved calculation, Eqs.~(\ref{eq:Ugamma})--(\ref{eq:Nsig}) are applied to each time interval, and $N_{\rm sig}^{\rm TR}$ is the sum of the event numbers obtained for the individual intervals. For the time-integrated calculation, the same equations are applied once to the spectrum fitted over the whole burst, which gives $N_{\rm sig}^{\rm TI}$. A parameter set is considered to be excluded when the expected number of events is large enough that at least one event would have been detected with $90\%$ probability, i.e., when $1-e^{-N_{\rm sig}}>0.9$ or, equivalently, $N_{\rm sig}>2.3$. The ratio $N_{\rm sig}^{\rm TR}/N_{\rm sig}^{\rm TI}$, which quantifies the difference between the two treatments, is shown in the right panels of Figures~\ref{Fig:1}-\ref{Fig:5}.
\bibliography{ms}{}
\bibliographystyle{aasjournalv7}

\begin{footnotesize}
\setlength{\tabcolsep}{4pt}
\renewcommand{\arraystretch}{0.85}
\begin{longtable*}{ccccccc}
\caption{Spectral parameters of Band function fits and inferred physical parameters for GRB 080916C} \\
\toprule
$t_1$ (s) & $t_2$ (s) & $\alpha$ & $E_{\rm p}$ (keV) & $\beta$ & $L_{\rm \gamma,\rm iso}$ (erg s$^{-1}$) & $E_{\rm \gamma,\rm iso}$ (erg) \\
\midrule
\endfirsthead
\multicolumn{7}{c}{{Continued from previous page}} \\
\toprule
$t_1$ (s) & $t_2$ (s) & $\alpha$ & $E_{\rm p}$ (keV) & $\beta$ & $L_{\rm \gamma,\rm iso}$ (erg s$^{-1}$) & $E_{\rm \gamma,\rm iso}$ (erg) \\
\midrule
\endhead
\bottomrule
\multicolumn{7}{r}{{Continued on next page}} \\
\endfoot
\bottomrule
\endlastfoot
0    & 3.7  & $-0.65 \pm 0.04$ & $488 \pm 45$   & $-2.64 \pm 0.15$ & $\sim \num{2.639953655309138e53}$ & $\sim \num{1.8257623410549178e53}$ \\
3.7  & 9.1  & $-1.10 \pm 0.02$ & $1342 \pm 214$ & $-2.26 \pm 0.04$ & $\sim \num{8.130567660353256e53}$ & $\sim \num{8.206554274001422e53}$ \\
9.1  & 17   & $-1.07 \pm 0.04$ & $664 \pm 135$  & $-2.18 \pm 0.04$ & $\sim \num{3.86693745030632e53}$ & $\sim \num{5.710057169611202e53}$ \\
17   & 25   & $-0.93 \pm 0.04$ & $420 \pm 59$   & $-2.18 \pm 0.03$ & $\sim \num{3.546289922420462e53}$ & $\sim \num{5.302863435395084e53}$ \\
25   & 41   & $-1.00 \pm 0.03$ & $512 \pm 68$   & $-2.30 \pm 0.04$ & $\sim \num{2.3152272779020274e53}$ & $\sim \num{6.924044195594849e53}$ \\
41   & 66   & $-0.91 \pm 0.05$ & $383 \pm 69$   & $-2.22 \pm 0.04$ & $\sim \num{1.1494742370331686e53}$ & $\sim \num{5.371374939407332e53}$ \\
\midrule
0    & 66   & $-0.96 \pm 0.02$ & $524 \pm 32$   & $-2.24 \pm 0.01$ & $\sim \num{2.727526199856861e53}$ & $\sim \num{3.3647986764589316e54}$
\label{table1}
\end{longtable*}
\end{footnotesize}

\begin{tiny}
\setlength{\tabcolsep}{2.5pt}
\renewcommand{\arraystretch}{0.8}
\begin{longtable*}{cccccccccc}
\caption{Spectral and derived physical parameters of GRB 110721A for Band and blackbody models}\\
\toprule
$t_1$ (s) & $t_2$ (s) & $T$ (keV) & $\alpha$ & $E_{\rm p}$ (keV) & $\beta$ & $E_{\rm \gamma,\rm iso,Band}$ (erg) & $E_{\rm \gamma,\rm iso,BB}$ (erg) & $\Gamma_{\rm ph}$ & $R_{\rm ph}$ (cm) \\
\midrule
\endfirsthead
\multicolumn{10}{c}{{Continued from previous page}} \\
\toprule
$t_1$ (s) & $t_2$ (s) & $T$ (keV) & $\alpha$ & $E_{\rm p}$ (keV) & $\beta$ & $E_{\rm \gamma,\rm iso,Band}$ (erg) & $E_{\rm \gamma,\rm iso,BB}$ (erg) & $\Gamma_{\rm ph}$ & $R_{\rm ph}$ (cm) \\
\midrule
\endhead
\bottomrule
\multicolumn{10}{r}{{Continued on next page}} \\
\endfoot
\bottomrule
\endlastfoot
-0.32 & 0.83 & $59 \pm 14$ & $-0.96 \pm 0.02$ & $6538 \pm 635$ & $-3.19 \pm 0.14$ & $\sim \num{1.0796766293365336e52}$ & $\sim \num{1.4677312476348136e50}$ & 394 & \num{6.472e11} \\
0.83 & 1.54 & $37 \pm 5$ & $-0.82 \pm 0.04$ & $1740 \pm 238$ & $-2.65 \pm 0.11$ & $\sim \num{5.133425862537453e51}$ & $\sim \num{1.5608825510937974e50}$ & 255 & \num{1.434e12} \\
1.54 & 2.18 & $34 \pm 4$ & $-0.98 \pm 0.05$ & $1109 \pm 237$ & $-2.89 \pm 0.23$ & $\sim \num{2.673074967011505e51}$ & $\sim \num{1.7879778975048862e50}$ & 201 & \num{1.514e12} \\
2.18 & 2.75 & $33 \pm 2$ & $-1.14 \pm 0.05$ & $1315 \pm 403$ & $-2.73 \pm 0.25$ & $\sim \num{2.1694483943014116e51}$ & $\sim \num{2.586747408557499e50}$ & 186 & \num{1.901e12} \\
2.75 & 3.46 & $21 \pm 3$ & $-1.14 \pm 0.06$ & $648 \pm 179$ & $-2.63 \pm 0.20$ & $\sim \num{1.7380376016403394e51}$ & $\sim \num{9.601156930098103e49}$ & 152 & \num{2.043e12} \\
3.46 & 4.35 & $12 \pm 3$ & $-1.03 \pm 0.10$ & $353 \pm 96$ & $-2.32 \pm 0.11$ & $\sim \num{1.431734005098957e51}$ & $\sim \num{4.519044129663019e49}$ & 123 & \num{2.896e12} \\
4.35 & 5.95 & $9 \pm 2$ & $-1.14 \pm 0.10$ & $528 \pm 146$ & $-2.46 \pm 0.11$ & $\sim \num{1.9148375831702765e51}$ & $\sim \num{3.2642120222621395e49}$ & 105 & \num{3.136e12} \\
5.95 & 8.38 & $5 \pm 2$ & $-1.03 \pm 0.15$ & $387 \pm 133$ & $-2.70 \pm 0.91$ & $\sim \num{1.5330763053684045e51}$ & $\sim \num{2.1557617123238888e49}$ & 74  & \num{4.147e12} \\
\midrule
-0.32 & 8.38 & $33 \pm 2$ & $-1.20 \pm 0.11$ & $2696 \pm 218$ & $-2.78 \pm 0.07$ & $\sim \num{2.9473845607958457e52}$ & $\sim \num{1.0789563697912848e51}$ & 217 & \num{1.102e12} 
\label{table2}
\end{longtable*}
\end{tiny}


\begin{tiny}
\setlength{\tabcolsep}{2pt}
\renewcommand{\arraystretch}{0.8}
\begin{longtable*}{cccccccc}
\caption{Spectral and derived physical parameters of GRB 090902B for blackbody and power-law model}\\
\toprule
$t_1$ (s) & $t_2$ (s) & $T$ (keV) & $\Gamma_{\rm PL}$ & $E_{\rm \gamma,\rm iso,BB}$ (erg) & $E_{\rm \gamma,\rm iso,PL}$ (erg) & $\Gamma_{\rm ph}$ & $R_{\rm ph}$ (cm) \\
\midrule
\endfirsthead
\multicolumn{8}{c}{{Continued from previous page}} \\
\toprule
$t_1$ (s) & $t_2$ (s) & $T$ (keV) & $\Gamma_{\rm PL}$ & $E_{\rm \gamma,\rm iso,BB}$ (erg) & $E_{\rm \gamma,\rm iso,PL}$ (erg) & $\Gamma_{\rm ph}$ & $R_{\rm ph}$ (cm) \\
\midrule
\endhead
\bottomrule
\multicolumn{8}{r}{{Continued on next page}} \\
\endfoot
\bottomrule
\endlastfoot

0    & 1.5  & $86 \pm 2$ & $-1.89 \pm 0.15$ & $\sim \num{4.907331806151426e52}$ & $\sim \num{7.901697087896601e51}$  & 522 & \num{2.244e12} \\
1.5  & 2.25 & $111 \pm 3$ & $-1.90 \pm 0.28$ & $\sim \num{3.700467199608738e52}$ & $\sim \num{2.6600691038196566e51}$ & 578 & \num{1.815e12} \\
2.25 & 2.81 & $132 \pm 4$ & $-1.87 \pm 0.28$ & $\sim \num{3.972296350085649e52}$ & $\sim \num{2.6020164524830093e51}$ & 666 & \num{1.765e12} \\
2.81 & 3.23 & $94 \pm 3$ & $-1.78 \pm 0.06$ & $\sim \num{2.0406602830692726e52}$ & $\sim \num{3.158728551569954e51}$  & 650 & \num{2.834e12} \\
3.23 & 3.83 & $114 \pm 3$ & $-1.84 \pm 0.09$ & $\sim \num{3.5780910934534637e52}$ & $\sim \num{3.6230454537456246e51}$ & 644 & \num{2.093e12} \\
3.83 & 4.46 & $103 \pm 3$ & $-1.82 \pm 0.06$ & $\sim \num{3.2302520729568524e52}$ & $\sim \num{4.074308361233653e51}$  & 627 & \num{2.321e12} \\
4.46 & 4.99 & $103 \pm 4$ & $-1.83 \pm 0.10$ & $\sim \num{2.100457706339302e52}$ & $\sim \num{3.7520890390457595e51}$ & 631 & \num{2.062e12} \\
4.99 & 5.45 & $124 \pm 4$ & $-1.87 \pm 0.33$ & $\sim \num{3.2121557783953396e52}$ & $\sim \num{1.9192732666701448e51}$ & 637 & \num{1.913e12} \\
5.45 & 5.86 & $134 \pm 4$ & $-1.83 \pm 0.10$ & $\sim \num{3.1057383318814534e52}$ & $\sim \num{2.3563199915311e51}$    & 704 & \num{1.892e12} \\
5.86 & 6.28 & $143 \pm 4$ & $-1.82 \pm 0.11$ & $\sim \num{4.832811623026081e52}$ & $\sim \num{2.4352900194909915e51}$ & 744 & \num{2.153e12} \\
6.28 & 6.61 & $165 \pm 5$ & $-1.83 \pm 0.11$ & $\sim \num{4.132050690884259e52}$ & $\sim \num{1.8821408154499006e51}$ & 795 & \num{1.798e12} \\
6.61 & 7.19 & $185 \pm 5$ & $-1.83 \pm 0.04$ & $\sim \num{7.993252296324609e52}$ & $\sim \num{9.978510941137618e51}$  & 946 & \num{1.796e12} \\
7.19 & 7.65 & $181 \pm 4$ & $-1.87 \pm 0.05$ & $\sim \num{7.687285160993066e52}$ & $\sim \num{9.696151799166229e51}$  & 922 & \num{2.002e12} \\
7.65 & 8    & $230 \pm 6$ & $-1.84 \pm 0.04$ & $\sim \num{8.580495284750167e52}$ & $\sim \num{1.335272511234267e52}$  & 1145 & \num{1.861e12} \\
8    & 8.5  & $211 \pm 5$ & $-1.87 \pm 0.02$ & $\sim \num{1.095054779912443e53}$ & $\sim \num{2.9228135312720937e52}$ & 1142 & \num{2.081e12} \\
8.5  & 9    & $212 \pm 5$ & $-1.86 \pm 0.02$ & $\sim \num{1.0545153747101866e53}$ & $\sim \num{3.040137766097297e52}$  & 1155 & \num{2.062e12} \\
9    & 9.5  & $213 \pm 5$ & $-1.89 \pm 0.02$ & $\sim \num{1.0161732832560687e53}$ & $\sim \num{3.289357280030001e52}$  & 1143 & \num{1.988e12} \\
9.5  & 10   & $149\pm 3$ & $-1.91 \pm 0.02$ & $\sim \num{8.269885890916558e52}$ & $\sim \num{3.26326423228204e52}$   & 947 & \num{3.022e12} \\
10   & 10.5 & $176 \pm 4$ & $-1.91 \pm 0.03$ & $\sim \num{8.929157468219842e52}$ & $\sim \num{2.2763928294930772e52}$ & 966 & \num{2.310e12} \\
10.5 & 11   & $195 \pm 5$ & $-1.87 \pm 0.03$ & $\sim \num{9.23683970370029e52}$ & $\sim \num{2.8672751942062507e52}$ & 1094 & \num{2.145e12} \\
11   & 11.5 & $152 \pm 3$ & $-1.91 \pm 0.03$ & $\sim \num{7.203183371010492e52}$ & $\sim \num{2.082290757204083e52}$  & 893 & \num{2.544e12} \\
11.5 & 12   & $154 \pm 4$ & $-1.91 \pm 0.03$ & $\sim \num{6.424029529854417e52}$ & $\sim \num{1.5656080851197033e52}$ & 863 & \num{2.266e12} \\
12   & 12.4 & $78 \pm 3$ & $-1.87 \pm 0.04$ & $\sim \num{1.6927919505496418e52}$ & $\sim \num{1.6391513696771286e52}$ & 719 & \num{4.235e12} \\
12.4 & 13.2 & $39 \pm 1$ & $-1.85 \pm 0.03$ & $\sim \num{1.923032609117693e52}$ & $\sim \num{3.456515576036325e52}$  & 560 & \num{1.003e13} \\
13.2 & 13.3 & $53 \pm 2$ & $-1.78 \pm 0.06$ & $\sim \num{8.28726501441397e51}$ & $\sim \num{5.2542948820621474e51}$ & 701 & \num{1.243e13} \\
13.3 & 13.6 & $50 \pm 2$ & $-1.81 \pm 0.04$ & $\sim \num{1.3963613046188718e52}$ & $\sim \num{1.6727973074066248e52}$ & 682 & \num{1.024e13} \\
13.6 & 13.8 & $59 \pm 2$ & $-1.84 \pm 0.07$ & $\sim \num{1.1397879974020219e52}$ & $\sim \num{7.596544318926961e51}$  & 625 & \num{7.413e12} \\
13.8 & 14.1 & $84 \pm 2$ & $-1.81 \pm 0.04$ & $\sim \num{2.5947082815932437e52}$ & $\sim \num{1.3552814116285025e52}$ & 791 & \num{5.811e12} \\
14.1 & 14.2 & $129 \pm 4$ & $-1.82 \pm 0.06$ & $\sim \num{2.0422196016557243e52}$ & $\sim \num{4.365428146200741e51}$  & 911 & \num{4.307e12} \\
14.2 & 14.4 & $124 \pm 3$ & $-1.82 \pm 0.06$ & $\sim \num{3.273923011475423e52}$ & $\sim \num{7.745752660377579e51}$  & 879 & \num{4.050e12} \\
14.4 & 14.6 & $78 \pm 3$ & $-1.81 \pm 0.04$ & $\sim \num{1.0115636177895434e52}$ & $\sim \num{8.757431501477828e51}$  & 794 & \num{5.087e12} \\
14.6 & 14.8 & $123 \pm 3$ & $-1.85 \pm 0.11$ & $\sim \num{2.8740562695266895e52}$ & $\sim \num{5.9211482209737165e51}$ & 810 & \num{3.517e12} \\
14.8 & 15   & $91 \pm 2$ & $-1.80 \pm 0.05$ & $\sim \num{2.1622328026129043e52}$ & $\sim \num{8.529205175611134e51}$  & 812 & \num{5.671e12} \\
15   & 15.1 & $125 \pm 4$ & $-1.78 \pm 0.11$ & $\sim \num{2.0538393901008678e52}$ & $\sim \num{4.1686155058256896e51}$ & 941 & \num{4.770e12} \\
15.1 & 15.2 & $117 \pm 3$ & $-1.79 \pm 0.13$ & $\sim \num{2.053232290935855e52}$ & $\sim \num{3.8825973078954636e51}$ & 891 & \num{5.126e12} \\
15.2 & 15.5 & $145 \pm 3$ & $-1.82 \pm 0.05$ & $\sim \num{5.90853699897659e52}$ & $\sim \num{9.782529459155931e51}$  & 919 & \num{3.373e12} \\
15.5 & 15.7 & $170 \pm 5$ & $-1.81 \pm 0.06$ & $\sim \num{4.307762017883987e52}$ & $\sim \num{6.51075250005026e51}$   & 1005 & \num{2.802e12} \\
15.7 & 16.2 & $84 \pm 2$ & $-1.82 \pm 0.03$ & $\sim \num{3.336631594449354e52}$ & $\sim \num{1.765508000037589e52}$  & 757 & \num{4.859e12} \\
16.2 & 16.3 & $112 \pm 4$ & $-1.84 \pm 0.13$ & $\sim \num{1.4944136254027387e52}$ & $\sim \num{3.337382595178742e51}$  & 793 & \num{4.260e12} \\
16.3 & 16.5 & $102 \pm 3$ & $-1.79 \pm 0.04$ & $\sim \num{2.6377395973524843e52}$ & $\sim \num{8.162201625298761e51}$  & 862 & \num{5.228e12} \\
16.5 & 16.7 & $109 \pm 3$ & $-1.81 \pm 0.06$ & $\sim \num{2.5596016130359966e52}$ & $\sim \num{7.278065547709854e51}$  & 836 & \num{4.368e12} \\
16.7 & 16.9 & $102 \pm 3$ & $-1.78 \pm 0.08$ & $\sim \num{1.8722496592059237e52}$ & $\sim \num{6.65165110536694e51}$   & 861 & \num{4.427e12} \\
16.9 & 17.1 & $53 \pm 3$ & $-1.84 \pm 0.06$ & $\sim \num{6.101105525437577e51}$ & $\sim \num{5.867406628778257e51}$  & 592 & \num{6.481e12} \\
17.1 & 17.5 & $68 \pm 2$ & $-1.83 \pm 0.04$ & $\sim \num{2.162097936319429e52}$ & $\sim \num{1.2181159747785876e52}$ & 660 & \num{5.766e12} \\
17.5 & 17.8 & $82 \pm 3$ & $-1.83 \pm 0.06$ & $\sim \num{1.4167067000788836e52}$ & $\sim \num{6.914854737189923e51}$  & 691 & \num{3.893e12} \\
17.8 & 18.3 & $53 \pm 2$ & $-1.83 \pm 0.04$ & $\sim \num{1.5939198969126108e52}$ & $\sim \num{1.333941000028523e52}$  & 591 & \num{6.619e12} \\
18.3 & 18.9 & $68 \pm 2$ & $-1.82 \pm 0.04$ & $\sim \num{2.435997506492839e52}$ & $\sim \num{1.9329727248612058e52}$ & 698 & \num{5.364e12} \\
18.9 & 19.4 & $61 \pm 2$ & $-1.79 \pm 0.04$ & $\sim \num{1.9381980117234915e52}$ & $\sim \num{1.5767986087540096e52}$ & 696 & \num{6.503e12} \\
19.4 & 19.6 & $55 \pm 2$ & $-1.81 \pm 0.08$ & $\sim \num{1.3341345747261732e52}$ & $\sim \num{7.135523943841835e51}$  & 632 & \num{9.348e12} \\
19.6 & 19.7 & $69 \pm 3$ & $-1.80 \pm 0.07$ & $\sim \num{8.392753068201054e51}$ & $\sim \num{4.30754283141483e51}$   & 732 & \num{7.659e12} \\
19.7 & 19.9 & $71 \pm 2$ & $-1.79 \pm 0.05$ & $\sim \num{1.6454796808768316e52}$ & $\sim \num{9.41450090836272e51}$   & 775 & \num{7.653e12} \\
19.9 & 20.1 & $88 \pm 3$ & $-1.80 \pm 0.07$ & $\sim \num{1.6238474103387318e52}$ & $\sim \num{9.210400344900074e51}$  & 834 & \num{5.295e12} \\
20.1 & 20.3 & $61 \pm 3$ & $-1.79 \pm 0.06$ & $\sim \num{7.585663371692221e51}$ & $\sim \num{7.769714662521409e51}$  & 734 & \num{6.694e12} \\
20.3 & 20.6 & $64 \pm 2$ & $-1.82 \pm 0.06$ & $\sim \num{1.4816344798088557e52}$ & $\sim \num{1.2143664874711882e52}$ & 690 & \num{6.613e12} \\
20.6 & 20.9 & $56 \pm 2$ & $-1.81 \pm 0.06$ & $\sim \num{1.3946162489775505e52}$ & $\sim \num{1.1449835736599942e52}$ & 659 & \num{7.847e12} \\
20.9 & 21   & $41 \pm 2$ & $-1.78 \pm 0.10$ & $\sim \num{4.097853108646847e51}$ & $\sim \num{3.5519397200545415e51}$ & 601 & \num{1.249e13} \\
21   & 21.3 & $35 \pm 2$ & $-1.82 \pm 0.06$ & $\sim \num{4.6295432360631894e51}$ & $\sim \num{9.348070832463198e51}$  & 544 & \num{9.754e12} \\
21.3 & 21.7 & $47 \pm 1$ & $-1.82 \pm 0.04$ & $\sim \num{1.535962377268234e52}$ & $\sim \num{1.180357573087807e52}$  & 578 & \num{8.770e12} \\
21.7 & 21.9 & $52 \pm 2$ & $-1.80 \pm 0.06$ & $\sim \num{6.68132653776853e51}$ & $\sim \num{7.840172298857562e51}$  & 681 & \num{8.007e12} \\
21.9 & 22.2 & $57 \pm 2$ & $-1.82 \pm 0.07$ & $\sim \num{1.08626055865511e52}$ & $\sim \num{9.900291620009424e51}$  & 651 & \num{6.605e12} \\
22.2 & 23   & $25 \pm 2$ & $-1.88 \pm 0.05$ & $\sim \num{2.6780138822193954e51}$ & $\sim \num{1.3064828210368588e52}$ & 423 & \num{6.868e12} \\
\midrule
0    & 23   & $117 \pm 1$ & $-1.83 \pm 0.01$ & $\sim \num{1.8487298628504422e54}$ & $\sim \num{7.665790952656914e53}$ & 851 & \num{3.075e12} \label{table3}
\end{longtable*}
\end{tiny}

\end{document}